\documentclass[10pt,twocolumn,twoside]{IEEEtran}

\IEEEoverridecommandlockouts

\usepackage{graphicx} 
\usepackage{epstopdf} 
\usepackage{cite}
\usepackage{amsmath,amssymb,amsfonts}
\usepackage{algorithmic}
\usepackage{graphicx}
\usepackage{textcomp}
\usepackage{xcolor}
\usepackage{color}

\usepackage{amssymb}
\usepackage{bbm} 
\usepackage{bm}

\usepackage{algorithm}
\usepackage{algorithmic}
\usepackage{url}
\usepackage{color}

\usepackage{graphicx} 
\usepackage{epstopdf} 

\usepackage{cite}
\usepackage{amsmath,amssymb,amsfonts}
\usepackage{algorithmic}
\usepackage{graphicx}
\usepackage{textcomp}
\usepackage{xcolor}
\usepackage{color}

\usepackage{amssymb}
\usepackage{bbm} 
\usepackage{bm}

\usepackage{algorithm}
\usepackage{algorithmic}
\usepackage{url}
\usepackage{color}

\usepackage{multirow} 

\usepackage{subfig}

\usepackage{tablefootnote}

\usepackage{caption}

\usepackage{amsmath}
\allowdisplaybreaks[1]

\usepackage{booktabs}

\def\BibTeX{{\rm B\kern-.05em{\sc i\kern-.025em b}\kern-.08em
		T\kern-.1667em\lower.7ex\hbox{E}\kern-.125emX}}
\begin{document}


\title{Privacy-aware VR streaming}
\author{\IEEEauthorblockN{Xing Wei and Chenyang Yang
	}
\\
	\IEEEauthorblockA{School of Electronics and Information Engineering, Beihang University, Beijing 100191, China\\ Email: \{weixing, cyyang\}@buaa.edu.cn}
}

\maketitle

\begin{abstract}
      Proactive tile-based virtual reality (VR) video streaming employs the current tracking data of a user to predict future requested tiles, then renders and delivers the predicted tiles to be requested before playback. The quality of experience (QoE) depends on the overall performance of prediction, computing (i.e., rendering) and communication. All prior works neglect that users may have privacy requirement, i.e., not all the current tracking data are allowed to be uploaded. In this paper, we investigate the privacy-aware VR streaming. We first establish a dataset that collects the privacy requirement of 66 users among 18 panoramic videos. The dataset shows that the privacy requirements of 360$^{\circ}$ videos are heterogeneous. Only 41\% of the total watched videos have no privacy requirement. Based on these findings, we formulate the privacy requirement as the \textit{degree of privacy} (DoP), and investigate the impact of DoP on the proactive VR streaming. First, we find that with DoP, the length of the observation window and prediction window of a tile predictor should be variable. Then, we jointly optimize the durations for computing and transmitting the selected tiles as well as the computing and communication capability, aimed at maximizing the QoE given arbitrary predictor and configured resources. From the obtained optimal closed-form solution, we find a resource-saturated region where DoP has no impact on the QoE and a resource-unsaturated region where the two-fold impacts of DoP are contradictory. On the one hand, the increase of DoP will degrade the prediction performance and thus degrade the QoE. On the other hand, the increase of DoP will improve the capability of computing and communication and thus improve the QoE. Simulation results using two predictors and a real dataset validate the analysis and demonstrate the overall impact of DoP on the QoE.
\end{abstract}



\maketitle

\section{Introduction}
Wireless virtual reality (VR) can provide a seamless and immersive experience to users. As the main type of VR services~, 360$^\circ$ video has the following unique features.
First, 360$^\circ$ video usually has $360^{\circ}\times 180^{\circ}$ panoramic view with ultra high resolution (e.g., binocular 16K \cite{HuaWei_Cloud_VR}). Second, the range of angles of a 360$^\circ$ video that humans can see at arbitrary time is only a small portion of the full panoramic view (e.g., 110$^\circ$ $\times$90$^\circ$), which is called field of view (FoV). More importantly, the stalls or black holes during watching 360$^\circ$ video will cause physiological discomfort, e.g., dizziness, which degrades the quality of experience (QoE) and thus should be avoid.

To stream such video with QoE guarantee, proactive VR video streaming is proposed \cite{optimizing_VR,adptive_tile}, which divides a full panoramic view segment into small tiles in spatial domain. Before the playback of the segment, the tiles to be most likely requested in the segment are first predicted using the user behavior-related data in an observation window, which are then computed (i.e., rendered) and finally delivered to the user.

While proactive VR video streaming is being intensive investigated in academia and industry, all the research neglects the willingness of users. \textit{Are users willing to share their behavior-related data while watching VR videos}?

Recent work shows that with less than 5 minutes tracking data while watching VR videos, simple prediction algorithm random forest can correctly identify 95\% of users among all the 511 users \cite{privacy_VR_identifiability}. This reveals that behavior-related data can be used to dig personal information. As the development of the technology, one may be able to dig more personal information than beyond imagination. 

With this consideration, it is reasonable to consider privacy protection in VR video streaming. Then, many fundamental problems arise: How to define the privacy requirement for VR video streaming? How to design privacy-aware VR video streaming? What is the impact of privacy requirement on the system? 

In this paper, we strive to answer these questions. The main contributions can be summarized as follows.

\begin{itemize}
	\item We establish the first privacy requirement dataset, from which we verify that users have heterogeneous privacy requirements.
	\item Based on the dataset, we define the degree of privacy (DoP) as the ratio of a video trace that a user requires not to disclose user behavior-related data.
	\item Based on the defined privacy requirement, we design a privacy-aware VR video streaming system. 
	We find a predictor with variable length of input and output is more suitable for privacy requirement. 
	Then, we optimize the durations and the capability for rendering and transmitting to maximize the QoE. 
	\item From the obtained optimal solution, we find a resource-saturated region where DoP has no impact and a resource-unsaturated region where DoP has contradictory impacts on the QoE. The increase of DoP on the one hand improves the capability of computing and communication and thus improve the QoE. On the other hand, it degrades the prediction performance and thus degrades the QoE. The overall impact depends on which factor dominates the QoE. 
	\item Simulation with two predictors on a real dataset verifies the analysis and shows that the impact of DoP on QoE also depends on other influence factors, e.g., the robustness of a predictor to the variation of DoP and assigned resources. The heterogeneous DoP calls for a predictor that has strong robustness to the variation of DoP.
\end{itemize}


\section{Privacy Investigation When Watching 360$^\circ$ Videos}

In this section, we investigate the privacy requirement of users when watching 360$^\circ$ videos. To this end, we first collect a dataset that contains the privacy requirement when watching 360$^{\circ}$ videos, then analyze the obtained data.

To obtain the real privacy requirement when a user requests a 360$^\circ$ video, the experiment is carefully designed to
mimic a real 360$^{\circ}$ video-on-demand (VoD) procedure. 

\subsection{Dataset Establishment}
The dataset includes the privacy requirement of 66 viewers when requesting three among 18 panoramic videos, called ``Privacy 360$^\circ$ video dataset". The age of viewers ranges from 18 to 38, most viewers are in their early twenties, and 40.9\% of them are female. The collection procedure is as follows.


First, we select 18 panoramic videos and one demo video from Bilibili\footnote{https://www.bilibili.com/} and Internet. Unlike other experiments that cutting videos into the video sequence \cite{NTHU_dataset} or removing the audio track \cite{xumai_predictHM}, we remain the complete videos in order to construct a real library of 360$^{\circ}$ videos. As shown in Table \ref{table:spe_360_video}, the content of these videos is diverse and covers the representative category of panoramic videos.

\begin{table*}[htbp]
	\caption{Description of 19 panoramic videos}\label{table:spe_360_video}
	\begin{center}
		\begin{tabular}{|c|c|c|c|}
			\hline
			\multicolumn{2}{|c|}{Category}&Video&Link\\
			\hline
			\multicolumn{2}{|c|}{Demo} &Guide Dog&www.bilibili.com/video/BV1gK411P7iB\\
			\hline
			\multirow{5}{*}{Landscape}&\multirow{3}{*}{Human landscape}&Venice carnival&www.bilibili.com/video/BV1hp4y1k79k\\
			\cline{3-4}
			&&Street concert&pan.baidu.com/s/190xMAwwy\_0EcLRl-kgRvSw, password:Stre\\
			\cline{3-4}
			&&Just Dance&pan.baidu.com/s/12raF1sPQNzrzbMPdKxVpuA, password:Just\\
			\cline{2-4}
			&\multirow{2}{*}{Natural landscape}&Lake baikal&www.bilibili.com/video/BV1b54y1C7W9\\
			\cline{3-4}
			&&Africa safari&pan.baidu.com/s/1YaPjPyBSzbcAxjkt3pkooA, password:Afri\\
			\cline{1-4}
			\multirow{4}{*}{Micro film}&\multirow{2}{*}{Science fiction}&Help&www.bilibili.com/video/BV1Wt4y1i7iH\\
			\cline{3-4}
			&&King Kong&www.bilibili.com/video/BV13f4y1q7CL\\
			\cline{2-4}
			&Horror&The Conjuring 2&www.bilibili.com/video/BV1Ua411w7Ut\\
			\cline{2-4}
			&Cartoon&Father and daughter&www.bilibili.com/video/BV1q5411576u\\
			\hline
			\multirow{3}{*}{Sports}&\multirow{2}{*}{First perspective}&Wingsuit flying&www.bilibili.com/video/BV1a64y1T7a4\\
			\cline{3-4}
			&&Val Thorens Ski&pan.baidu.com/s/1HWyYfRmt35eiVrLNqkV9Jw, password:ValT\\
			\cline{2-4}
			&Third perspective&Basketball&pan.baidu.com/s/1mgNIzkOjghli3Lw3hRa8QA, password:Bask\\
			\cline{1-4}
			\multicolumn{2}{|c|}{Life sharing}&Travel dream imagine&pan.baidu.com/s/1tZVqbb1Z1sgy6XrncCEDuw, password:Trav\\
			\hline
			\multicolumn{2}{|c|}{\multirow{2}{*}{Work of art}}&Dreams of Dali&www.bilibili.com/video/BV14D4y1d7Gp\\
			\cline{3-4}
			\multicolumn{2}{|c|}{}&Zero-gravity in dreams&www.bilibili.com/video/BV1ap4y1Y7YU\\
			\hline
			\multicolumn{2}{|c|}{Videos of game}&Assassin's Creed&www.bilibili.com/video/BV1QA411j7hw\\
			\hline
			\multicolumn{2}{|c|}{\multirow{2}{*}{First perspective interaction}}&WannaOne's girlfriend&www.bilibili.com/video/BV1Bv41167dM\\
			\cline{3-4}
			\multicolumn{2}{|c|}{}&Surrounded by Korean girls&www.bilibili.com/video/BV1cV411y7oR\\
			\hline
		\end{tabular}
	\end{center}
\end{table*}

Before the video requests start, viewers are guided to wear the HMD of an HTC Vive Pro, seat on a swivel chair, and interact using a handheld controller. They can turn around freely, such that all panoramic regions are accessible. Then they are guided to watch a demo video, which helps them to adapt to the virtual reality environment.
When viewers have acclimatized themselves to the VR environment, the video requests start, viewers can freely choose three panoramic videos they prefer. Unlike other experiments where viewers are forced to watch all the videos \cite{xumai_predictHM,NTHU_dataset, dataset_2}, the free choice mechanism avoids viewers to watch the videos they dislike, thus the obtained privacy requirement of videos can be more close to reality.
After watching the 360$^{\circ}$ videos, they are asked to fill out a questionnaire, as shown in Table \ref{tab:question}.
\begin{table*}[htbp]   
	\caption{\label{tab:question}Questionnaire}   
	\begin{tabular}{l}    
		\toprule
Please fill out the privacy requirement of the three videos (0$\sim$100\%). 1. first video: \_\_\_\_, 2. second video: \_\_\_\_, 3. third video: \_\_\_\_\\ 
		\midrule 
		\texttt{Explanation}: (a) From the viewing orientations trace of a video one can recover the image in FoVs that you choose to see.\\
		(b) Privacy requirement is the percentage of the trace that you are not willing to share with others.\\
		(c) Three examples of privacy requirement: 0: Share the complete trace to communication operator or online video content\\ platform, 100\%: Not share any part of the traces. 50\%: Share half length to the operator and platform, keep half by yourself.\\
		\bottomrule\\   
\end{tabular} 
\end{table*}

\subsection{Results analysis}

\begin{figure}[htbp]
	\centering
	\begin{minipage}[c]{0.85\linewidth}
		\centering
		\includegraphics[width=1\textwidth]{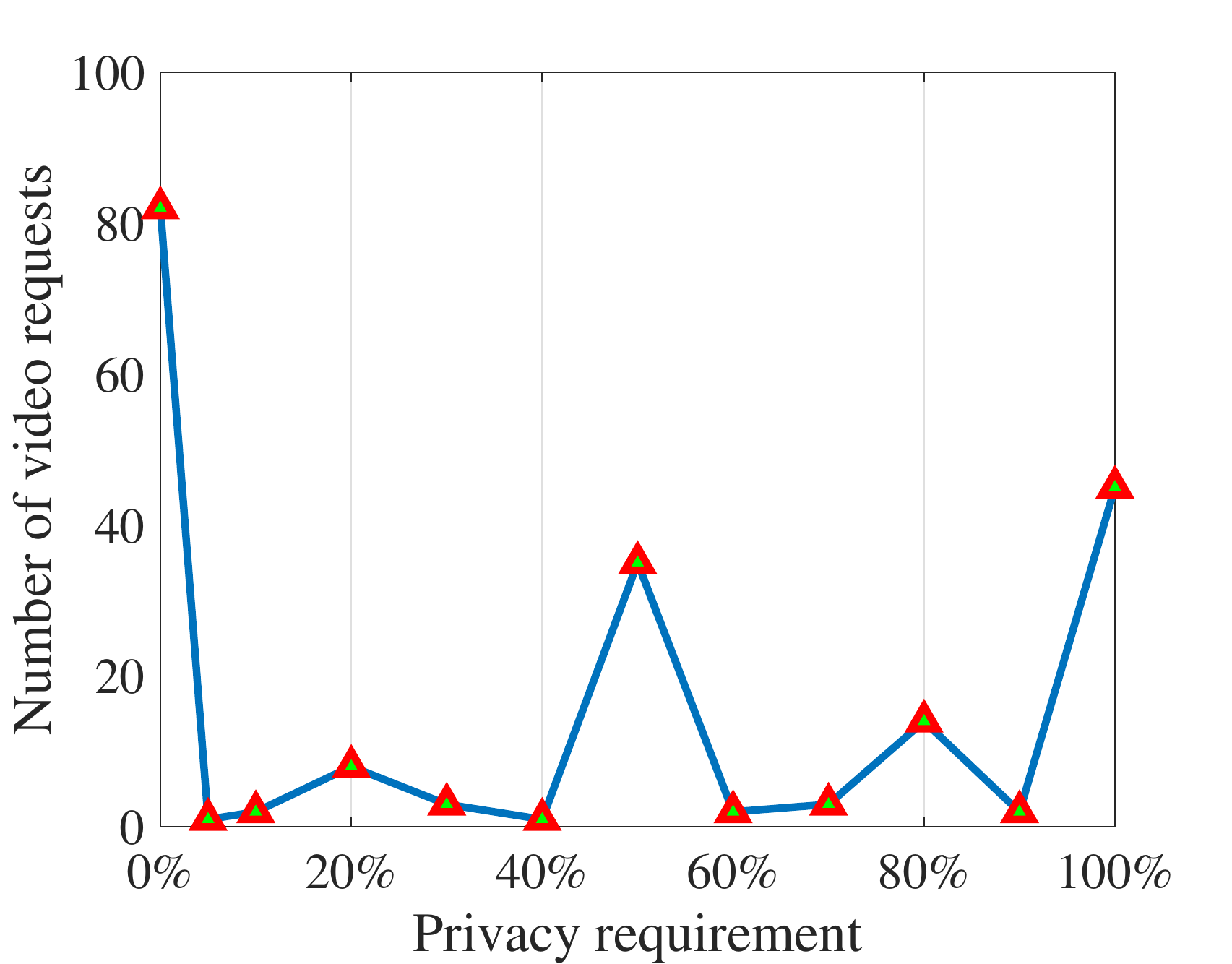}
	\end{minipage}
	\caption{Privacy requirement and corresponding ratio}\label{Fig:dataset_statsi}	
\end{figure}

\begin{figure}[htbp]
	\centering
	\begin{minipage}[c]{0.85\linewidth}
		\centering
		\includegraphics[width=1\textwidth]{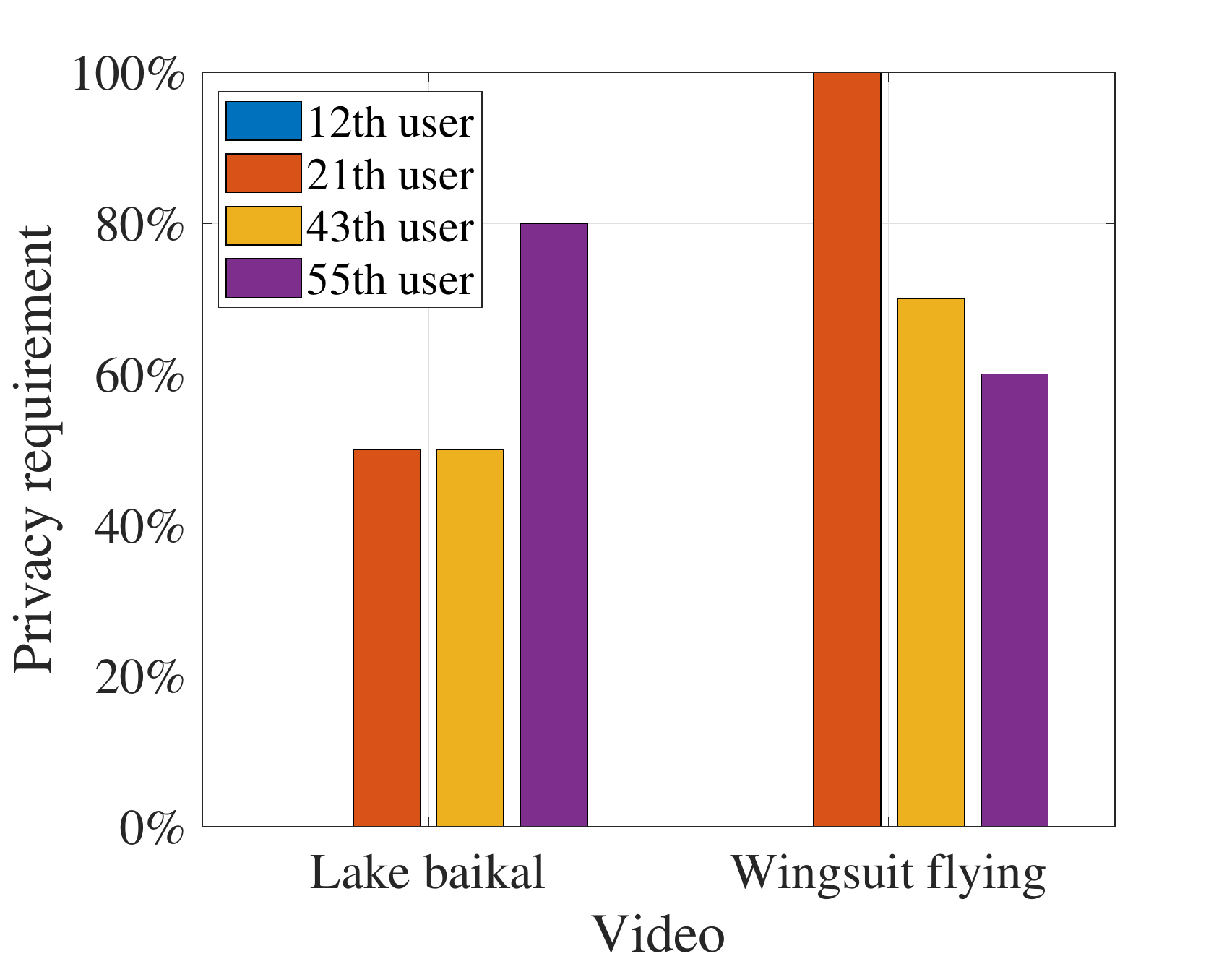}
	\end{minipage}
	\caption{Heterogeneous privacy requirements among users and videos}\label{Fig:het_privacy_req}	
\end{figure}

We collect 3$\times$66 = 198 privacy requirements, the statistical result is shown in Figure \ref{Fig:dataset_statsi}. Only $\frac{81}{198}=41\%$ of the total video requests has no privacy requirement, i.e., privacy requirement equals to 0. Furthermore, from Figure \ref{Fig:het_privacy_req}, users have heterogeneous privacy requirement of 360$^{\circ}$ videos. From these results we can find that privacy requirements are necessary to be considered in proactive VR video streaming system. Therefore, in the following, we consider the privacy-aware VR video streaming system and investigate the impact of privacy requirement on the system.

\section{System Model}\label{section-system_model}
Consider a tile-based proactive VR video streaming system with an multi-access edge computing (MEC) server co-located with a base station (BS) that serves $K$ users.
The MEC server equips with powerful computing units for rendering and training predictors, and can access a VR video library, which can be realized by local caching or high speed backhaul hence the delay from the Internet to the MEC server can be omitted.
Each user requests 360$^{\circ}$ videos from the library according to their own time schedule and interests hence the requests are asynchronous, which then can be decoupled into multiple individual requests. In the sequel, without loss of generality, we consider arbitrary one request for $v$th video from the $k$th user for analysis.

The playback duration of VR video is $T_{\textit{VR}}$. Each VR video consists of $L$ segments in temporal domain, and each segment consists of $M$ tiles in spatial domain. The playback duration of each tile equals to the playback duration of a segment, denoted by $T_{\mathrm{seg}}$ \cite{optimizing_VR,survey_Hsu}.

Each user is equipped with a HMD, which can measure user behavior-related data\footnote{The current useful user behavior-related data for prediction include sensor-related data, (i.e., the head movement orientation tracking  \cite{optimizing_VR,verylong_predict} and eye tracking data \cite{xumai_predictHM,vive} from the HMD sensors) and content-related data (i.e., the images in historical FoVs \cite{xumai_predictHM,NTHU_dataset}, temporal-spatial saliency \cite{Fixation_Prediction,CVPR_gaze}, and audio track information \cite{audio_prediction}).
While \textit{our framework is applicable for prediction using one or multiple types of data}, we take the head movement trace as an example for easy exposition.
}, send the recorded data to the MEC server, and pre-buffer segments. 


\subsection{Degree of Privacy}

To reflect the privacy requirement when watching 360$^{\circ}$ videos, we define the \textit{degree of privacy} as the percentage of the ``private duration" among the playback duration of a 360$^{\circ}$ video, i.e.,
\begin{equation}\label{DoP_def}
	\rho_{} \triangleq \frac{T_{\textit{VR}}^p}{T_{\textit{VR}}}\in[0,100\%]
\end{equation}
where $T_{\textit{VR}}^p$ is the total private duration in the VR video for an arbitrary user, wherein the user requires not to disclose any user behavior-related data.
When $\rho_{}=0$, the user allows to share all the data during watching the VR video. When $\rho=100\%$, the user forbids to release any data during the whole playback procedure. The DoP of the $k$th user when requesting the $v$th video can be expressed as $\rho_{k,v}$.

The total available duration where the user behavior-related data is accessible by the MEC server is $T_{\textit{VR}}^a = T_{\textit{VR}} - T_{\textit{VR}}^p = (1 - \rho_{})\cdot T_{\textit{VR}}$.
In order to maximally utilize the available data with duration $T_{\textit{VR}}^a$ for predicting tile requests in each segment, we assume the duration $T_{\textit{VR}}^a$ is equally allocated among all the segments.
%
Then, for each segment, the duration of the available data used for prediction, i.e., the observation window, is
\begin{align}\label{T_obw_def}
	t_{\mathrm{obw}}(\rho_{}) = \frac{T_{\textit{VR}}^a}{L} =  (1 - \rho_{})\cdot \frac{ T_{\textit{VR}}}{L}  = (1 - \rho_{})\cdot T_{\mathrm{seg}}
\end{align}
Then, the private duration in each segment is 
\begin{align*}
	t_{\mathrm{pv}}(\rho_{}) = T_{\mathrm{seg}} - t_{\mathrm{obw}}(\rho_{}) = \rho T_{\mathrm{seg}}
\end{align*}

\begin{figure}[htbp]
	\centering
	\begin{minipage}[t]{1\linewidth}
		\includegraphics[width=1\textwidth]{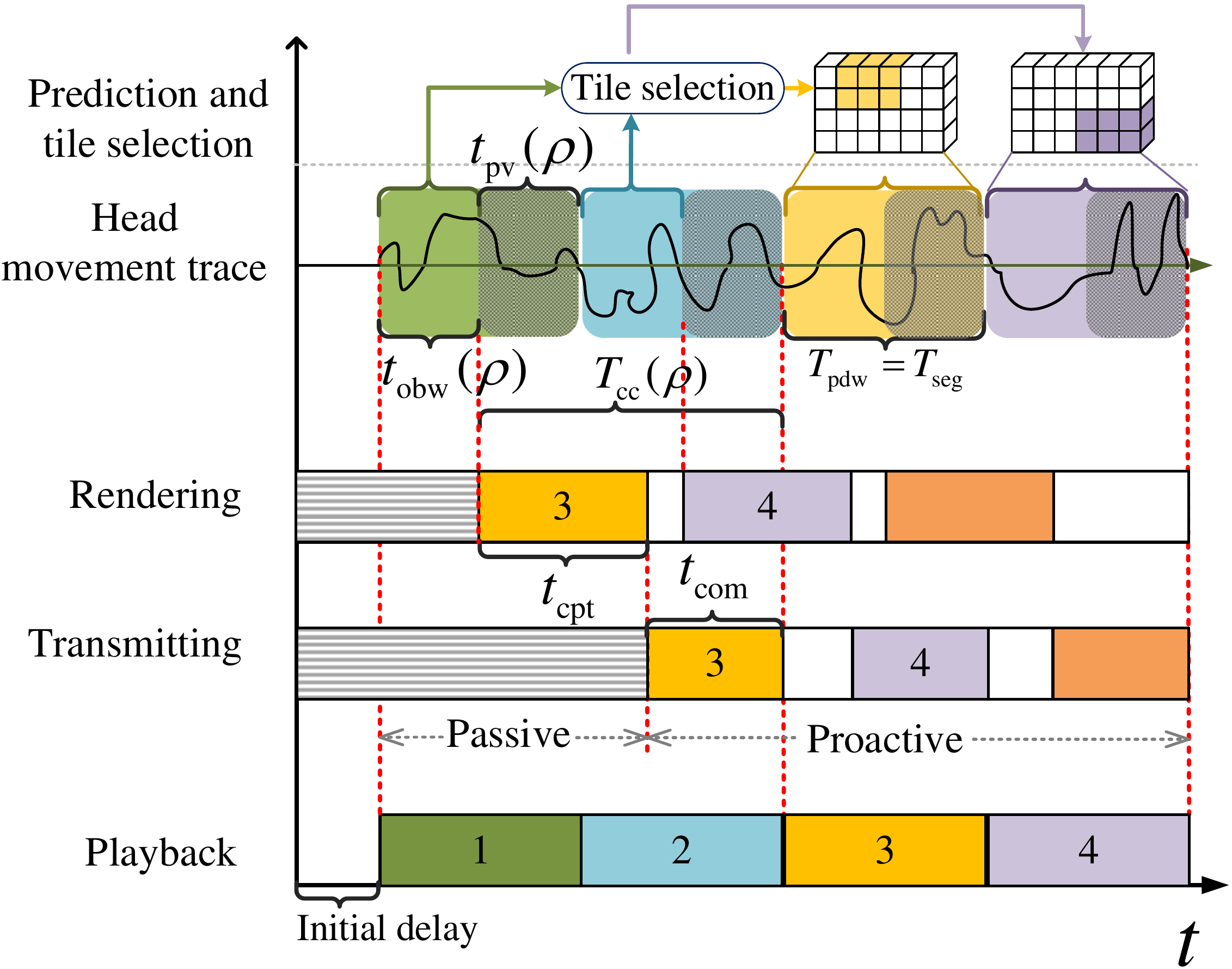}
	\end{minipage}
	\caption{Streaming the first four segments of a VR video, $\rho>0$.}
	\label{Fig:MEC4VR_pipeline}
\end{figure}

\subsection{Privacy-aware Streaming Procedure}

When a user requests a VR video with DoP $\rho>0$, the MEC server first streams the initial ($l_{0}-1$)th segments in a passive streaming mode \cite{transmission_mode-standard-update}. When the MEC server collects the user behaviour-related data (e.g., the head movement trace) in an observation windows with duration $t_{\mathrm{obw}}(\rho)$, the tiles to be played in the $l_0$th segment with duration $T_{\mathrm{pdw}}=T_{\mathrm{seg}}$ can be predicted. Then, proactive streaming for the $l_{0}$th segment begins, subsequent segments are predicted, rendered, and transmitted one after another, as shown in Figure 
\ref{Fig:MEC4VR_pipeline}. In the figure, $l_{0}=3$.

Specifically, at the end of the observation window for the first segment, tile request probabilities or the fixation sequences of FoVs in the third (i.e., the $l_0$th) segment can be predicted.\footnote{Predicting the second segment may be useless. For example, when $\rho=0$, there is no time for proactively rendering and transmitting the second segment.} 
Based on the probabilities (or the fixation sequences) and the number of tiles that can be streamed, the tiles to be streamed can be determined (to be explained in Section \ref{section:tile_selection}).
Then, the selected tiles are rendered with duration $t_{\mathrm{cpt}}$ and the sequence of FoVs can be generated, and finally the rendered tiles are transmitted with duration $t_{\mathrm{com}}$, which should be finished before the start time of playback for the predicted segment. The total duration for computing and communication is
\begin{align*}
	T_{\mathrm{cc}}(\rho)=T_{\mathrm{pv}}(\rho) + T_{\mathrm{seg}} = (1+\rho)T_{\mathrm{seg}}
\end{align*}
which begins from the end of the observation window and terminates at the playback time of the predicted segment.
The durations for computing and transmitting satisfy 
\begin{align*}
	t_{\mathrm{com}}(\rho) + t_{\mathrm{cpt}}(\rho)= (1+\rho)T_{\mathrm{seg}}
\end{align*}
which depends on $\rho$, and thereby are functions of $\rho$.


\subsection{Computing and Transmission Model}


The computing resource of MEC for rendering a VR video can be assigned by allocating graphics processing unit (GPU) and compute unified device architecture cores \cite{Nvidia_cloudXR,Xing_VR_Shannon}.
The configured computing resource for rendering a video from the request of the $k$th user is denoted as $\mathcal{F}_{\mathrm{cpt},k}$ (in floating-point operations per second, FLOPS). Then, the number of bits that can be rendered per second, referred to as the \textit{computing rate}, is
\begin{displaymath}
	C_{\mathrm{cpt},k} \triangleq \frac{\mathcal{F}_{\mathrm{cpt},k}}{\mu_r} (\textit{in bit/s}),
\end{displaymath}
where $\mu_r$ is the required floating-point operations (FLOPs) for rendering one bit of FoV \textit{in FLOPs/bit} \cite{Xing_VR_Shannon}.

The BS serves $K$ single-antenna users using zero-forcing beamforming with $N_t$ antennas.
The instantaneous data rate at the $i$th time slot within the duration $t_{\mathrm{com}}$ for the $k$th user is
\begin{align*}
	C_{\mathrm{com},k}^{i}(B_k, p_k)=B_k\log_2 \left(1+\frac{p_k d_k^{-\alpha}|\tilde{h}^i_k|^2}{\sigma^2} \right),
\end{align*}
where $B_k$ is the configured bandwidth for the $k$th user, $\tilde{h}^i_k\triangleq (\mathbf{h}^i_k)^H\mathbf{w}^i_k$ is the equivalent channel gain, $p_k$ and $\mathbf{w}^i_k$ are respectively the transmit power and beamforming vector for the $k$th user,  $d_k$ and $\mathbf{h}^i_k\in\mathbb{C}^{N_t}$ are respectively the distance and the small scale channel vector from the BS to the $k$th user, $\alpha$ is the path-loss exponent, $\sigma^2$ is the noise power, and $(\cdot)^{H}$ denotes conjugate transpose.

We consider indoor users as in the literature, where the distances of users, $d_k$, usually change slightly \cite{LSTM_update,survey_Hsu,NTHU_dataset} and hence are assumed fixed.
Due to the head movement and the variation of the environment,
small-scale channels are time-varying, which are assumed as remaining constant in each time slot with duration $\Delta T$ and changing independently with identical distribution among time slots. With the proactive transmission, the rendered tiles (i.e., sequences of FoVs) in a segment should be transmitted with duration $t_{\mathrm{com}}(\rho)$. The number of bits transmitted with $t_{\mathrm{com}}(\rho)$ can be expressed as $\overline{C}_{\mathrm{com},k}(B_k, p_k)\cdot t_{\mathrm{com}}(\rho)$, where 
\begin{align*}
	\overline{C}_{\mathrm{com},k}(B_k, p_k) \triangleq \frac{1}{N_s}\sum_{i=1}^{N_s}C_{\mathrm{com},k}^{i}(B_k, p_k) \cdot \Delta T
\end{align*}
is the time average transmission rate, and $N_s$ is the number of time slots in $t_{\mathrm{com}}$.
Since future channels are unknown when making the optimization, we use ensemble-average rate $\mathbbm{E}_h\{C_{\mathrm{com},k}(B_k, p_k)\}$ \cite{ergodic-capacity} to approximate the time-average rate $\overline{C}_{\mathrm{com},k}(B_k, p_k)$, where $\mathbbm{E}_h\{\cdot\}$ is the expectation over $h$, which can be very accurate when $N_s$ or $N_t$ is large \cite{Xing_VR_Shannon}. 

We can observe that the ensemble-average transmission rate $\mathbbm{E}_h\{C_{\mathrm{com},k}(B_k, p_k)\}$ can be configured by allocating bandwidth $B_k$ and transmission power $p_k$. In the sequel, we consider arbitrary one user and use $C_{\mathrm{com}}$ and $C_{\mathrm{cpt}}$ to replace $\mathbbm{E}_h\{C_{\mathrm{com},k}(B_k, p_k)\}$ and $C_{\mathrm{cpt}}$ for notional simplicity. Besides, we refer the transmission rate and computing rate as ``\textit{Resource rates}".


\section{Problem Formulation}

%

\subsection{Performance Metric of Tile Prediction}\label{section:DoO_def}

Average segment degree of overlap (average-DoO) has been used to measure the prediction performance for a VR video \cite{Xing_VR_Shannon}. It indicates the average overlap of the predicted tiles and the requested tiles among all the proactively streamed segments , which is defined as
\begin{align*}
	\mathcal{D}(t_{\mathrm{obw}}) \triangleq \frac{1}{L-l_0 + 1}\sum_{l=l_0}^L\frac{\mathbf{q}_{l}^\mathsf{T}\cdot\mathbf{e}_{l}({t_{\mathrm{obw}}})  }{\|\mathbf{q}_{l}\|_1}\in[0,100\%]
\end{align*}
where $\mathbf{q}_{l}\triangleq [q_{l,1},...,q_{l,M}]^\mathsf{T}$ denotes the ground-truth of the tile requests for the $l$th segment with $q_{l,m}\in\{0,1\}$, $\mathbf{e}_{l}({t_{\mathrm{obw}}})\triangleq [e_{l,1}({t_{\mathrm{obw}}}),...,e_{l,M}({t_{\mathrm{obw}}})]^\mathsf{T}$ denotes the predicted tile requests for the segment with $e_{l,m}({t_{\mathrm{obw}}})\in\{0,1\}$,
$(\cdot)^\mathsf{T}$ denotes transpose of a vector, and $\|\cdot\|_1$ denotes the $\ell_1$ norm of a vector. When the $m$th tile in the $l$th segment is truly requested, $q_{l,m}=1$, otherwise $q_{l,m}=0$. When the tile is predicted to be requested, $e_{l,m}({t_{\mathrm{obw}}})=1$, otherwise it is zero. 

If the number of predicted tiles $\|\mathbf{e}_{l}\left(t_{\mathrm{obw}}\right)\|_1$ increases, average-DoO will be nondecreasing. 
We consider $\|\mathbf{e}_{l}\left(t_{\mathrm{obw}}\right)\|_1 = N_{\textit{fov}}$ with $N_{\textit{fov}} < M$, where $N_{\textit{fov}}$ is the number of tiles within a FoV.
A larger value of average-DoO indicates a better prediction.

As the verified Assumption 1 in \cite{Xing_VR_Shannon} states, a predictor can be more accurate with a longer observation window. Therefore, average-DoO is a monotonically increasing function of $t_{\mathrm{obw}}$.
When consider DoP, $t_{\mathrm{obw}}$ becomes a monotonically decreasing function of $\rho_{}$, as shown in \eqref{T_obw_def}. Then, average-DoO becomes a monotonically decreasing function of $\rho_{}$ as
\begin{align}\label{seg_DoO}
	\mathcal{D}_{}(\rho_{})\triangleq\mathrm{DoO}\left(t_{\mathrm{obw}}(\rho)\right),  \ \ \frac{\partial\mathcal{D}}{\partial\rho} < 0
\end{align}
From here on we omit the intermediate variable $t_{\mathrm{obw}}$, all the prior functions of $t_{\mathrm{obw}}$ are expressed as the functions of $\rho$, e.g., $\mathbf{e}_{l}({t_{\mathrm{obw}}})$ can be expressed as $\mathbf{e}_{l}(\rho)$.

\subsection{Computing and Communication Capability}
To reflect the ratio of tiles in a segment that can be rendered and transmitted within assigned durations for transmission and computing, define the capability of computing and communication(CC) tasks as

\begin{align}
	C_{\mathrm{cc}} \triangleq \frac{N(\rho)}{M}\in[0,1], \label{def:C_cc}
\end{align}
where
\begin{align*}
	&N(\rho) \triangleq \min\left\{\frac{C_{\mathrm{com}}t_{\mathrm{com}}(\rho)}{s_{\mathrm{com}}^{}}, \frac{C_{\mathrm{cpt}}t_{\mathrm{cpt}}(\rho)}{s_{\mathrm{cpt}}^{}}, M\right\}
\end{align*}
is the number of tiles that can be computed in $t_{\mathrm{cpt}}$ and transmitted in $t_{\mathrm{com}}$, $s_{\mathrm{com}} = {px}_w \cdot{px}_h \cdot b \cdot r_f \cdot T_{\mathrm{seg}}/\gamma_{c}$ \cite{HuaWei_Cloud_VR} is the number of bits in each tile for transmission, $s_{\mathrm{cpt}} = {px}_w \cdot{px}_h \cdot b \cdot r_f \cdot T_{\mathrm{seg}}$ is the number of bits in a tile for rendering, ${px}_w$ and ${px}_h$ are the pixels in wide and high of a tile, $b$ is the number of bits per pixel relevant to color depth \cite{HuaWei_Cloud_VR}, $r_f$ is the frame rate, and $\gamma_c$ is the compression ratio.

\subsection{Determine the streamed tiles}\label{section:tile_selection}

When the CC capability ($N(\rho)$) is determined, the streamed tiles can be selected by either one of the following schemes. (a) When the predicted parameter is the probabilities of the tile requests \cite{Fixation_Prediction}, the first $N(\rho)$ tiles with largest probabilities are selected. (b) When the predicted parameter is the fixation sequence of FoVs \cite{optimizing_VR,Romero_ACM_MMsys_20_paper}, $N(\rho)$ can be first transformed into the size of an ellipse, then the selected tiles is the tiles within the ellipse \cite{FoV_edge_streaming}.

\subsection{Metric of Quality of Experience}
For proactive tile-based streaming, the motion-to-photon latency can be zero if the constraint $t_{\mathrm{com}}(\rho) + t_{\mathrm{cpt}}(\rho) = T_{\mathrm{cc}}(\rho)$ can be satisfied.
Yet black holes will appear if the requested tiles cannot be streamed before playback.
For arbitrary given predictor, DoP $\rho_{}$, and the number of selected tiles $N(\rho)$, we consider the following QoE metric
\begin{align}\label{QoE_cal}
	\mathrm{QoE} &\triangleq \frac{1}{L-l_0 + 1}\sum_{l=l_0}^L\frac{\mathbf{q}_{l}^\mathsf{T}\cdot\mathbf{e}_{l}(\rho)  }{\|\mathbf{q}_{l}\|_1},
\end{align}
which is the percentage of the correctly streamed tiles that can be computed and delivered among all the requested tiles,
where $\|\mathbf{e}_{l}\left(\rho\right)\|_1 = N(\rho) = C_{\mathrm{cc}}(\rho)\cdot M$.
We can observe that the QoE depends on the average-DoO and CC capability, which can be expressed as
\begin{align}
	\mathrm{QoE} = \mathcal{Q}\left(\mathcal{D}\left(\rho_{}\right), C_{\mathrm{cc}}(\rho)\right)\in[0,100\%]\label{QoE-Metric}
\end{align}
When all the requested tiles in a VR video are proactively computed and delivered before playback, the value of the QoE is $100\%$. 
When $C_{\mathrm{cc}}(\rho)$ is increased, more tiles can be rendered and transmitted, then more requested tiles can be satisfied. Therefore, we have the following remark:

\textbf{Remark 1}: $\mathcal{Q}\left(\mathcal{D}\left(\rho_{}\right), C_{\mathrm{cc}}(\rho)\right)$ is a monotonically nondecreasing function of $C_{\mathrm{cc}}(\rho)$.

\section{Privacy-Aware Prediction: Variable Length of Input and Output}
\begin{figure}[htbp]
	\centering
	\subfloat[Prediction window for training and testing.]{\label{Fig:variable_input_output_train_test}
		\begin{minipage}[c]{1\linewidth}
			\centering
			\includegraphics[width=1\textwidth]{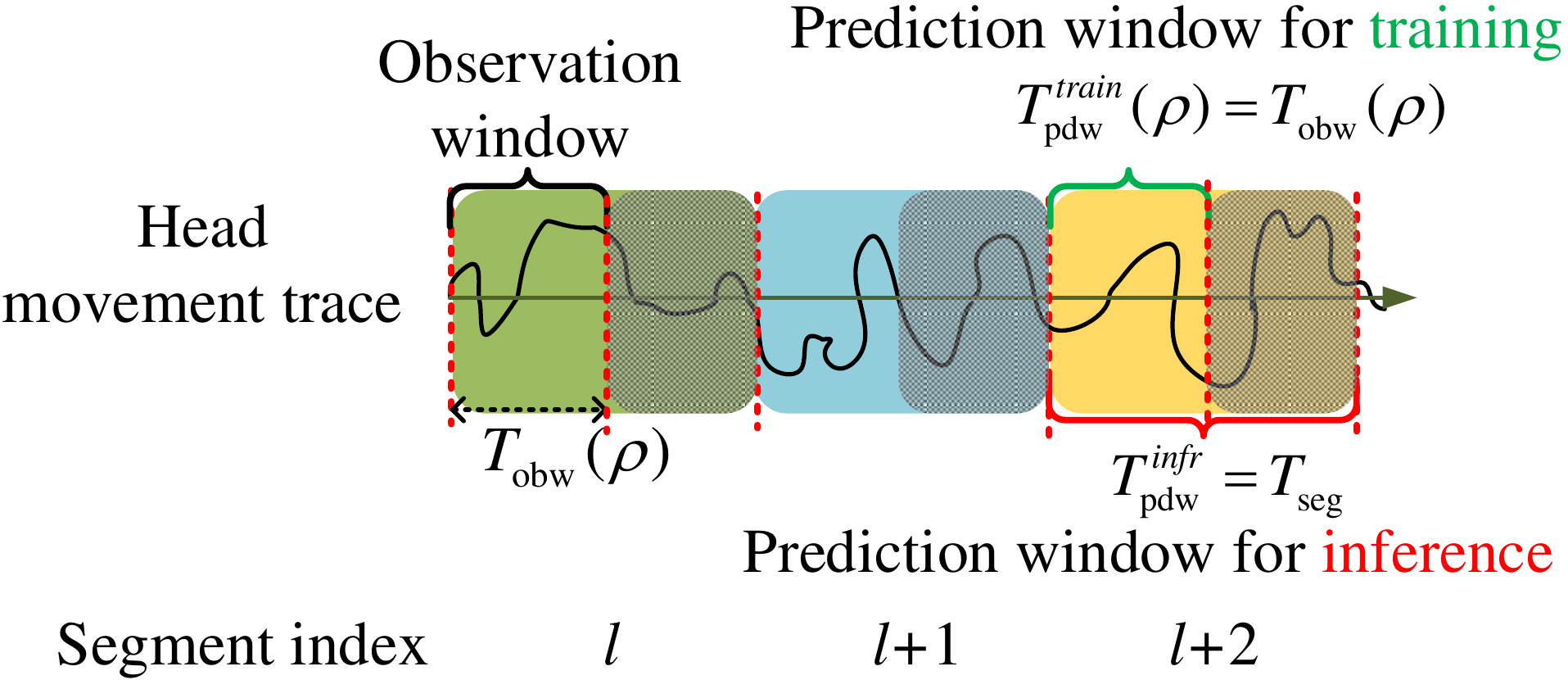}
		\end{minipage}
	}\\
	\subfloat[Training using the data of $k_1$th user for $v$th video, $\rho_{k_1,v}=50\%$.]{\label{Fig:variable_input_output_user1}
		\begin{minipage}[c]{1\linewidth}
			\centering
			\includegraphics[width=1\textwidth]{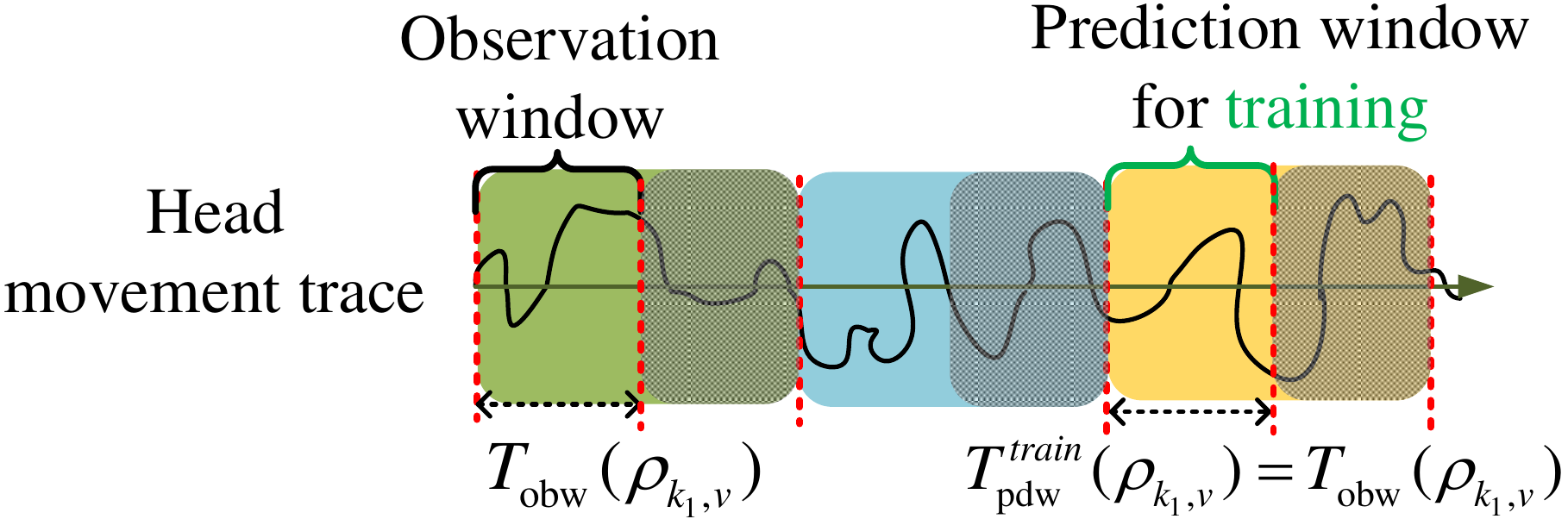}
		\end{minipage}
	}\\
	\subfloat[Training using the data of $k_2$th user for $v$th video, $\rho_{k_2,v}=80\%$.]{\label{Fig:variable_input_output_user2}
		\begin{minipage}[c]{1\linewidth}
			\centering
			\includegraphics[width=1\textwidth]{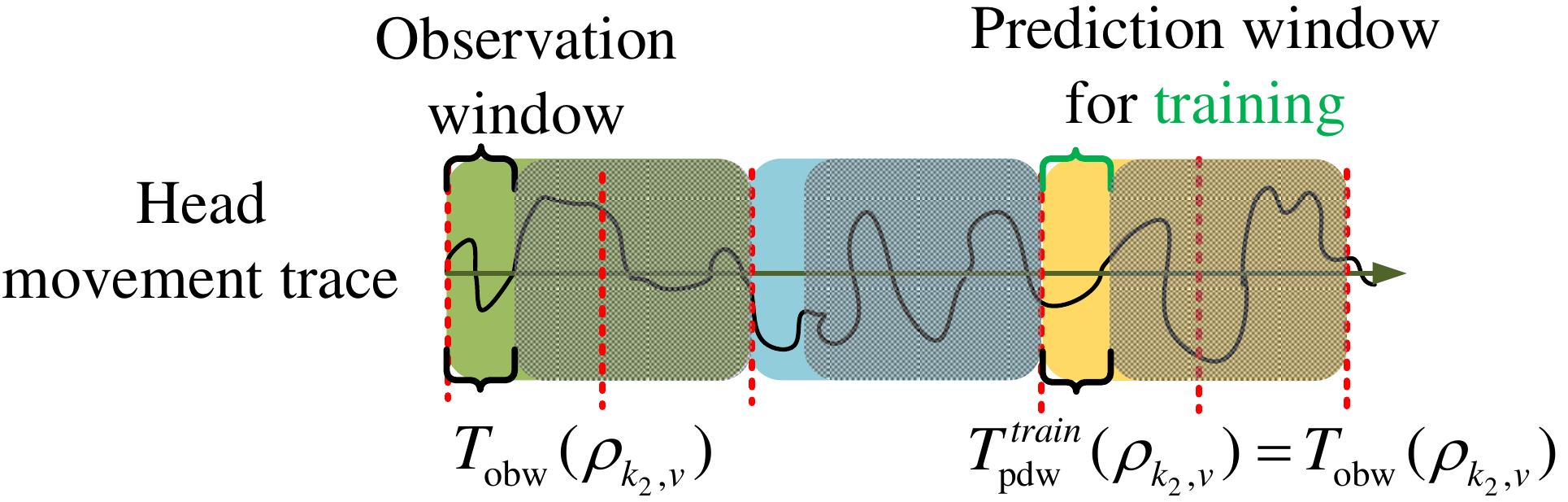}
		\end{minipage}
	}
	\caption{Observation and prediction windows for training and inference}\label{Fig:pos_only_baselines}	
\end{figure}

Without consideration of privacy, to train the predictor and predict requested tiles, the durations of observation and prediction windows are fixed \cite{Romero_ACM_MMsys_20_paper}. 
However, when users have privacy requirement, i.e., $\rho>0$, the durations of these windows should be \textit{variable}. The reasons are two-folds.
(1) The durations of prediction window in training and inference are different, as shown in Figure \ref{Fig:variable_input_output_train_test}. To train a predictor and learn the mapping from data in $T_{\mathrm{obw}}(\rho)$ of the $l$th segment to the tile request probabilities of the ($l$+2)th segment, the prediction window in the ($l$+2)th segment should be $T_{\mathrm{seg}}$. However, the accessible data in the ($l$+2)th segment used as labels is with duration $T_{\mathrm{obw}}(\rho)$. Therefore, the duration of prediction window for training becomes $T_{\mathrm{pdw}}^{\textit{train}}(\rho)=T_{\mathrm{obw}}(\rho)$. When predicting the tiles, the predictor needs to infer the tile requests in a whole segment, i.e., the prediction window for inference should be $T_{\mathrm{pdw}}^{\textit{infr}}=T_{\mathrm{seg}}$. (2) In training procedure, the durations of observation and prediction windows among different users and videos may be variable, as shown in \ref{Fig:variable_input_output_user1} and \ref{Fig:variable_input_output_user2}. This is because users may have heterogeneous privacy requirements.

To handle the variable length of observation and prediction windows caused by privacy requirement, 
a predictor with dimensional scalability of input and output could be more suitable, e.g., the sequence-to-sequence framework \cite{seq2seq}. 

\section{DoP: Contradictory Effects for QoE}
In this section, we investigate how DoP affects the QoE. To this end, we first optimize durations for communication and computing as well as the CC capability to maximize the QoE. From the obtained closed-form solution, 
we find a resource-saturated and a resource-unsaturated regions. Then we investigate how the DoP affects average-DoO and CC capability in the two regions, respectively. Finally, we discuss the overall impact of DoP on the QoE.
 
\subsection{Joint Optimization of the Durations and Capability for Computing and Communication}
Given arbitrary computing rate $C_{\mathrm{cpt}}$, transmission rate $C_{\mathrm{com}}$ and DoP $\rho$, we find the optimal durations for communication and computing as well as the CC capability to achieve maximized QoE,
i.e.,
\begin{subequations}
	\begin{align*}
		\textbf{P0}:& \max_{t_{\mathrm{com}}(\rho),t_{\mathrm{cpt}}(\rho), C_{\mathrm{cc}}(\rho)} \  \ \mathcal{Q}\left(\mathcal{D}\left(\rho_{}\right), C_{\mathrm{cc}}(\rho)\right) \\
		&  \ \ \ \  \  s.t.  \ \    C_{\mathrm{cc}}(\rho) = \frac{\min\left\{\frac{C_{\mathrm{com}}t_{\mathrm{com}}(\rho)}{s_{\mathrm{com}}^{}}, \frac{C_{\mathrm{cpt}}t_{\mathrm{cpt}}(\rho)}{s_{\mathrm{cpt}}^{}}, M\right\}}{M}, \\
		& \ \ \ \ \  \  \ \  \ \ \ t_{\mathrm{com}}(\rho) + t_{\mathrm{cpt}}(\rho) = (1+\rho)T_{\mathrm{seg}},
	\end{align*}
\end{subequations}
Problem \textbf{P0} is a functional optimization problem, where the optimal solution as a function of $C_{\mathrm{com}}$, $C_{\mathrm{cpt}}$, and $\rho$ needs to obtained. To emphasize the impact of $\rho$, we do not explicitly express $t_{\mathrm{com}}(\rho)$, $t_{\mathrm{cpt}}(\rho)$, and $C_{\mathrm{cc}}(\rho)$ as functions of $C_{\mathrm{com}}$ and $C_{\mathrm{cpt}}$. 
As derived in the supplementary material, the solution of \textbf{P0} is,
\begin{subequations}\label{P3:opt_solution_general}
	\begin{align}
		&t_{\mathrm{com}}^{*}(\rho) \!=\!\!
		\left
		\{\begin{array}{lr}
			\frac{C_{\mathrm{cpt}}s_{\mathrm{com}}(1+\rho)T_{\mathrm{seg}}}{C_{\mathrm{com}}s_{\mathrm{cpt}} + C_{\mathrm{cpt}}s_{\mathrm{com}}},&\!\!\!\! (1+\rho)T_{\mathrm{seg}} \leq T_{\mathrm{cc}}^M,  \\
			\left[\frac{s_{\mathrm{com}}M}{C_{\mathrm{com}}}, (1+\rho)T_{\mathrm{seg}} - \frac{s_{\mathrm{cpt}}M}{C_{\mathrm{cpt}}}\right], & \!\!\!\!(1+\rho)T_{\mathrm{seg}} \geq T_{\mathrm{cc}}^M,\\
		\end{array}
		\right.\label{P3:opt_solution_general_t_cpt}\\
			&t_{\mathrm{cpt}}^{*}(\rho) \!=\!\!
		\left
		\{\begin{array}{lr}
			\frac{C_{\mathrm{cpt}}s_{\mathrm{com}}(1+\rho)T_{\mathrm{seg}}}{C_{\mathrm{com}}s_{\mathrm{cpt}} + C_{\mathrm{cpt}}s_{\mathrm{com}}},&\!\!\!\! (1+\rho)T_{\mathrm{seg}} \leq T_{\mathrm{cc}}^M,  \\
			\left[\frac{s_{\mathrm{cpt}}M}{C_{\mathrm{cpt}}}, (1+\rho)T_{\mathrm{seg}} - \frac{s_{\mathrm{com}}M}{C_{\mathrm{com}}}\right], & \!\!\!\!(1+\rho)T_{\mathrm{seg}} \geq T_{\mathrm{cc}}^M,\\
		\end{array}
		\right.
		\label{P3:opt_solution_general_t_com}\\
		& C_{\mathrm{cc}}^*(\rho) =
		\left
		\{\begin{array}{lr}
			\frac{(1+\rho)T_{\mathrm{seg}}}{T_{\mathrm{cc}}^{M}} ,& (1+\rho)T_{\mathrm{seg}} \leq T_{\mathrm{cc}}^M,  \\
			1,& (1+\rho)T_{\mathrm{seg}} \geq T_{\mathrm{cc}}^M,
		\end{array}
		\right.\label{P3:opt_solution_general_N}
	\end{align}
\end{subequations}
where $T_{\mathrm{cc}}^{M}$ is the duration to deliver and compute all the $M$ tiles in a segment with expression
\begin{align}\label{T_CC_max}
	T_{\mathrm{cc}}^{M} \triangleq \frac{s_{\mathrm{com}}M}{C_{\mathrm{com}}} + \frac{s_{\mathrm{cpt}}M}{C_{\mathrm{cpt}}}.
\end{align}
The value of $1/{T_{\mathrm{cc}}^{M}}$  monotonically increases with the increase of $C_{\mathrm{com}}$ and $C_{\mathrm{cpt}}$, which can reflect the resource rates.


\subsection{Resource-saturated and Resource-Unsaturated Regions}
In the following, we show that the system may operate in a resource-saturated or a resource-unsaturated region.
From \eqref{P3:opt_solution_general} we can find that the optimal solution can be divided into two regions with a boundary line $(1+\rho)T_{\mathrm{seg}}=T_{\mathrm{cc}}^M$.

When $(1+\rho)T_{\mathrm{seg}}\geq T_{\mathrm{cc}}^M$, the duration for computing and transmitting the streamed tiles, i.e., $(1+\rho)T_{\mathrm{seg}}$, is no less than the required duration for computing and transmitting all the tiles $T_{\mathrm{cc}}^M$. This indicates that all the tiles can be computed and transmitted, i.e., $C_{\mathrm{cc}}^*(\rho)=1$, as shown in \eqref{P3:opt_solution_general_N}. Besides, since all tiles can be streamed when the configured resource rates (reflected by $1/T_{\mathrm{cc}}^M$) achieves $(1+\rho)T_{\mathrm{seg}} = T_{\mathrm{cc}}^M$, further increasing the resource rates  is useless for improving the maximized CC capability $C_{\mathrm{cc}}^*(\rho)$. That is to say, the resource rates are \textit{saturated} in this region. 
We refer to this region as ``\textit{Resource-saturated region}". 

When $(1+\rho)T_{\mathrm{seg}}< T_{\mathrm{cc}}^M$, the duration for computing and transmitting the selected tiles is less than the required duration for computing and transmitting all the tiles. This indicates that not all the tiles can be computed and transmitted. In this region, increasing the resource rates (i.e., $1/T_{\mathrm{cc}}^M$) can always increase the percentage of rendered and transmitted tiles, i.e., $C_{\mathrm{cc}}^*(\rho)=\frac{(1+\rho)T_{\mathrm{seg}}}{T_{\mathrm{cc}}^{M}}$. We refer to this region as ``\textit{Resource-unsaturated region}".

In the resource-saturated region, $C_{\mathrm{cc}}^*(\rho)=1$, all tiles can be streamed before playback, then QoE = $100\%$. In this region, DoP has no impact on the QoE. In the following, we discuss the impact of DoP in the resource-unsaturated region.
%
%

%
%
%
\subsection{Contradictory Role in the Resource-Unsaturated Region}
\subsubsection{Improve the Maximized CC Capability}
The maximized CC capability indicates the maximal percentage of streamed tiles among all the tiles with duration $T_{\mathrm{cc}}(\rho)$, which can be regarded as ``completed amount" of CC tasks with given resource rates and time. We can rewrite the maximized CC capability in the resource-unsaturated region from \eqref{P3:opt_solution_general_N} as follows:
\begin{align*}
	C_{\mathrm{cc}}^*(\rho) = \underbrace{1/{T_{\mathrm{cc}}^M}}_{\textit{resource \ rates\ gain}}  \cdot   \underbrace{(1+\rho)T_{\mathrm{seg}}}_{\textit{duration gain}},
\end{align*}
The first term corresponds to a $\textit{resource rates gain}$. The second term corresponds to a $\textit{duration gain}$, which comes from the fact that the increase of $\rho$ increases the duration for computing and communication $T_{\mathrm{cc}}(\rho)$. We can observe that the increase of $\rho$ can be equivalently transformed into the increase of resource rates for improving the CC capability. Besides, from \eqref{T_CC_max} we can find that either increasing the transmission rate $C_{\mathrm{com}}$ or the computing rate $C_{\mathrm{cpt}}$ can provide resource rates gain. This indicates that \textit{different combinations of computing and transmission rates can achieve the same resource rates gain}.

\begin{figure}[htbp]
	\centering
	\begin{minipage}[t]{0.85\linewidth}
		\includegraphics[width=1\textwidth]{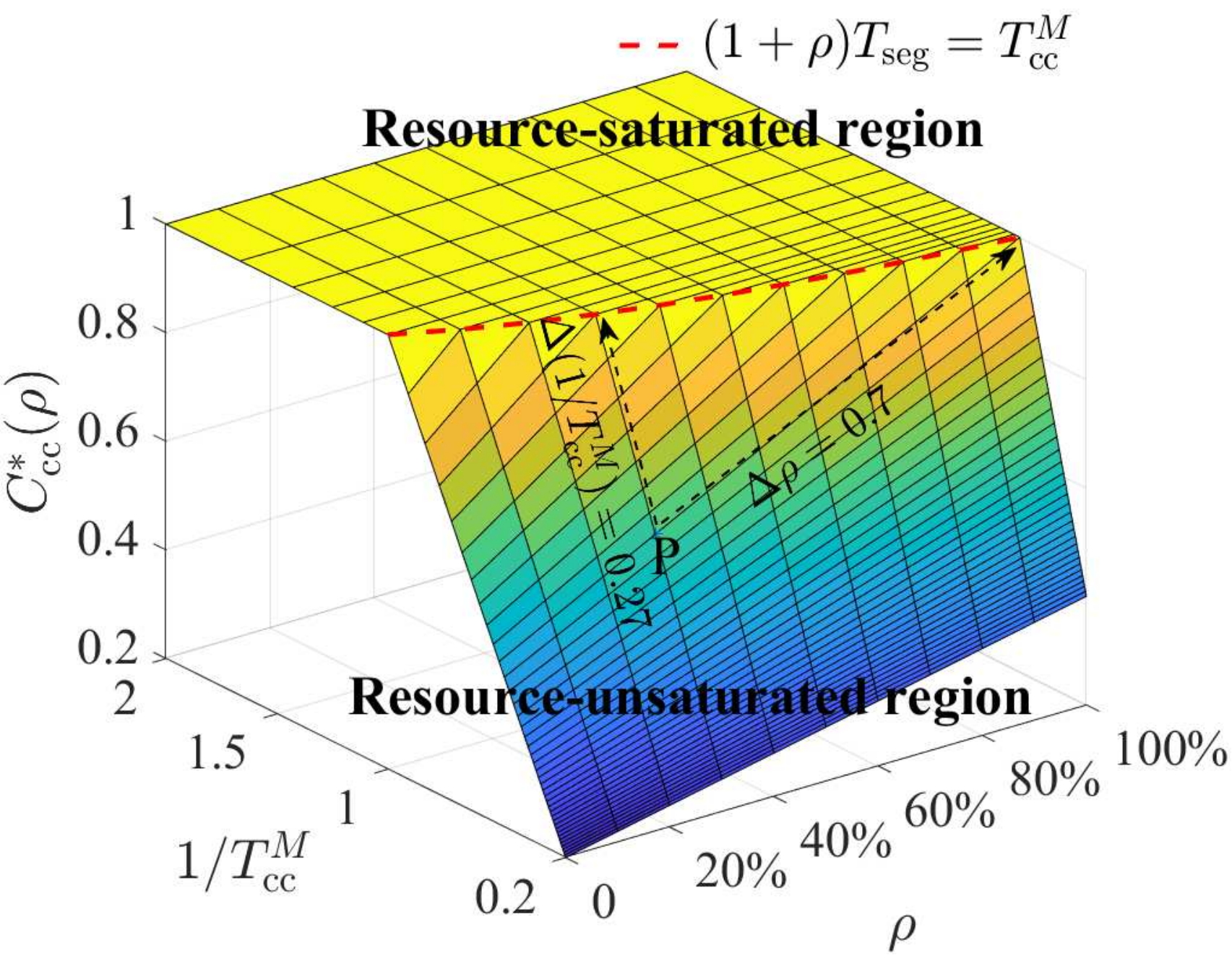}
	\end{minipage}
	\caption{Resource-saturated and resource-unsaturated regions.}
	\label{Fig:Resource-saturated_Region}
\end{figure}

To visualize the impact of $\rho$ in the two regions in term of CC capability, in Figure \ref{Fig:Resource-saturated_Region} we provide values of $C_{\mathrm{cc}}^*(\rho)$ obtained from \eqref{P3:opt_solution_general_N}. Specifically, consider the point ``P" in the resource-unsaturated region. To achieve the resource-saturated region, we can see that increasing the DoP by 0.7 is equivalent to increasing $1/T_{\mathrm{cc}}^M$ by 0.27.

\subsubsection{Degrade the Average-DoO}

From \eqref{seg_DoO}, we can see that the increase of DoP decreases the duration for observation and then degrades the average-DoO. 



\subsubsection{Overall Impact} 
In the resource-unsaturated region, the impact of DoP is complicated. On the one hand, it improves the CC capability. On the other hand, it degrades the average-DoO. Generality speaking, the overall impact depends on if the QoE is dominated by the increment of CC capability or the reduction of average-DoO.  

\section{Simulation and Numerical Results}
In this section, we show the overall impact of DoP on QoE via simulation and numerical results. 

First we consider the prediction task on a real dataset \cite{NTHU_dataset}, where 300 traces of head movement positions from 30 users\footnote{The original dataset contains the traces of 50 users. According to the analysis in  \cite{Romero_ACM_MMsys_20_code,Romero_ACM_MMsys_20_paper}, the traces of the first 20 users may have some mistakes, thus we only use the traces of the other 30 users.} watching 10 VR videos are used for training and testing predictors. The ratio of training and testing sets is 8:2.  
We use two predictors, \texttt{position-only} and \texttt{no-motion} predictors, which achieve the state-of-the-art accuracy in the dataset, according to tests in \cite{Romero_ACM_MMsys_20_code}. The \texttt{position-only} predictor employs a sequence-to-sequence LSTM-based architecture, which only exploits the time series of past head movement positions as input, to predict the time series of future positions \cite{Romero_ACM_MMsys_20_code}. Note that the predictor does not consider the time required for computing and communication as well as the degree of privacy. To reserve time for computing and communication and satisfying the DoP requirement, we tailor the predictor as follows. Set the distance between the observation and prediction windows as $T_{\mathrm{cc}}(\rho)$, set the durations of observation window, prediction window for training, prediction window for inference (i.e., testing) as $T_{\mathrm{obw}}(\rho), T_{\mathrm{obw}}(\rho), T_{\mathrm{seg}}$, respectively, as shown in Figure \ref{Fig:variable_input_output_train_test}. We refer to the predictor as \texttt{tailored position-only} predictor. Other details and hyper-parameters of the tailored predictor is the same as the \texttt{position-only} predictor \cite{Romero_ACM_MMsys_20_code}. The \texttt{no-motion} predictor assumes no head movement and simply uses the last position in the observation window as the predicted time series of future positions \cite{Romero_ACM_MMsys_20_code}.

Since one value of $1/T_{\mathrm{cc}}^M$ can represent various combinations of computing and transmission rates, we directly set the range of $1/T_{\mathrm{cc}}^M$ to reflect the variation of resources rates. The procedure to obtain the mapping from the configured resources to $1/T_{\mathrm{cc}}^M$ can be found in the supplementary material, for the interested readers.


The playback duration of a segment is set as $T_{\mathrm{seg}}=1$ s \cite{FoV_aware_tile}. The total number of tiles is $M=200$ \cite{NTHU_dataset}. The size of FoVs are modeled by $100^{\circ}\times100^{\circ}$ circles \cite{FoV_size_NTHU,Romero_ACM_MMsys_20_code}. According to our tests, the average number of tiles in a FoV is $N_{\textit{fov}}=33$. 
To gain useful insight, we assume that all users have identical DoP requirement for all videos, ranging from 0 to 80\%. Given DoP and $1/T_{\mathrm{cc}}^M$, the maximized CC capability $C_{\mathrm{cc}}^*(\rho)$ can be obtained from \eqref{P3:opt_solution_general_N}. Then, given $C_{\mathrm{cc}}^*(\rho)$ and predicted time series of positions, the streamed tiles can be determined by scheme (b) in Section \ref{section:tile_selection}. Finally, QoE is first calculated from \eqref{QoE_cal}, then averaged over the testing set.

\begin{figure}[htbp]
	\centering
	\subfloat[\texttt{No-motion} predictor]{\label{Fig:qoe_rho_T_cc_m_No_motion}
		\begin{minipage}[c]{0.85\linewidth}
			\centering
			\includegraphics[width=1\textwidth]{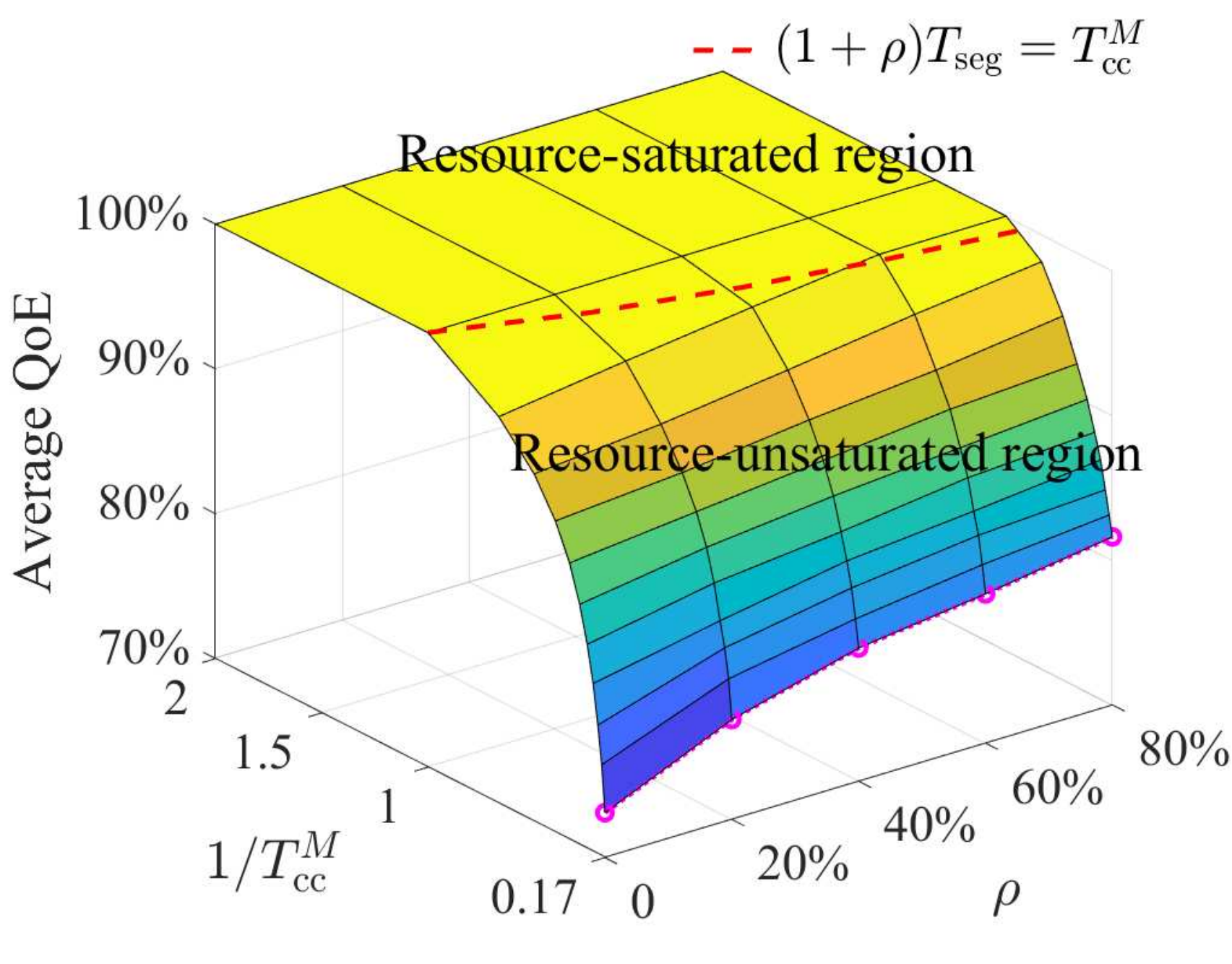}
		\end{minipage}
	}\\
	\subfloat[\texttt{Tailored position-only} predictor]{\label{Fig:qoe_rho_T_cc_m_pos_only}
		\begin{minipage}[c]{0.85\linewidth}
			\centering
			\includegraphics[width=1\textwidth]{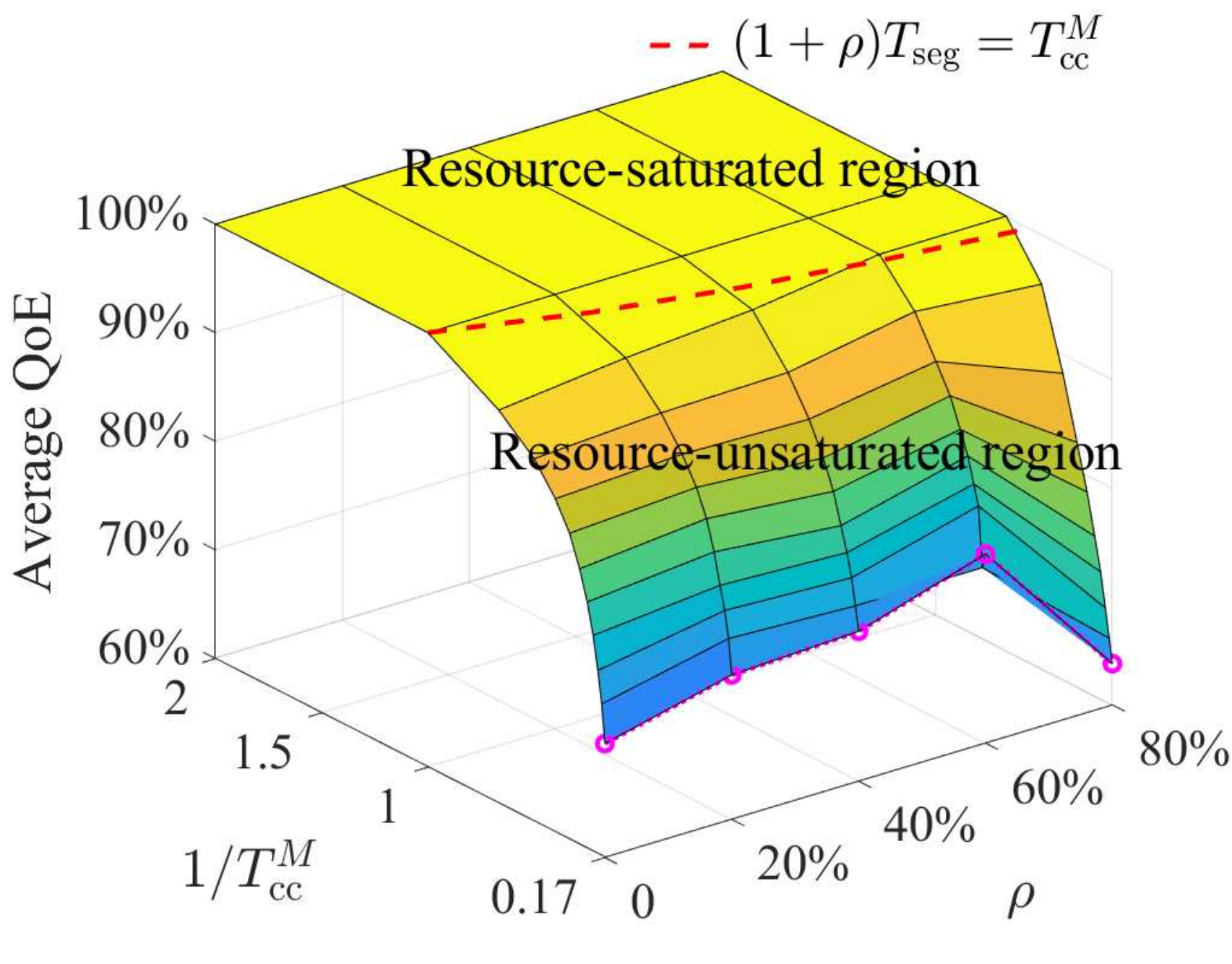}
		\end{minipage}
	}
	\caption{Average QoE v.s. resource rates and DoP}\label{Fig:qoe_rho_T_cc_m}
\end{figure}

In Figure \ref{Fig:qoe_rho_T_cc_m}, we show the average QoE achieved by two predictors versus DoP and the assigned resource rates. 
We can observe that the impact of DoP in the resource-unsaturated region depends on the predictor and the assigned resource rates. For the \texttt{no-motion} predictor, as shown in Figure \ref{Fig:qoe_rho_T_cc_m_No_motion}, regardless of the assigned resource rates, the increase of DoP slightly improves the QoE. For the \texttt{tailored position-only} predictor, as shown in Figure \ref{Fig:qoe_rho_T_cc_m_pos_only}, when assign high resource rates, the impact of DoP is not obvious, when assign less resource rates, the impact of DoP is more complicated. 

To further understand how the final QoE is affected by the DoP and why the impact also depends on the predictor in the resource-unsaturated region, we consider one case when the resource rates $1/T_{\mathrm{cc}}^M=0.17$ as an example, to investigate how the CC capability $C_{\mathrm{cc}}^*(\rho)$, average-DoO $\mathcal{D}(\rho)$, and average QoE is effected by DoP $\rho$.
As shown in
Figure \ref{Fig:QoE, CC capability, and DoO v.s. DoP}. For the \texttt{no-motion} predictor, we observe that the increase of DoP increases the maximized CC capability $C_{\mathrm{cc}}^*(\rho)$ and degrades the average-DoO $\mathcal{D}(\rho)$. Since the degradation of average-DoO is relative small, the QoE is dominated by the increase of CC capability. For the \texttt{tailored position-only} predictor, the variation of average-DoO is relative high, then the QoE is dominated by the variation of average-DoO. Besides, we can find that for different predictors, the sensitivity of average-DoO in terms of DoP is different. When a predictor is more sensitive, e.g., \texttt{tailored position-only} predictor, the QoE is more likely dominated by the average-DoO. When a predictor is more robust to the variation of DoP, e.g., the \texttt{no-motion} predictor, the QoE is more likely dominated by the variation of CC capability. This is the reason why the impact of DoP also depends on the predictor. In practice, the DoP are heterogeneous rather than identical. This 
calls for predictors that have strong robustness to the variation of DoP.

\begin{figure}[htbp]
	\centering
	\subfloat[No-motion predictor]{\label{Fig:qoe_rho_C_cc_No_motion_T_cc_m_6}
		\begin{minipage}[c]{0.85\linewidth}
			\centering
			\includegraphics[width=1\textwidth]{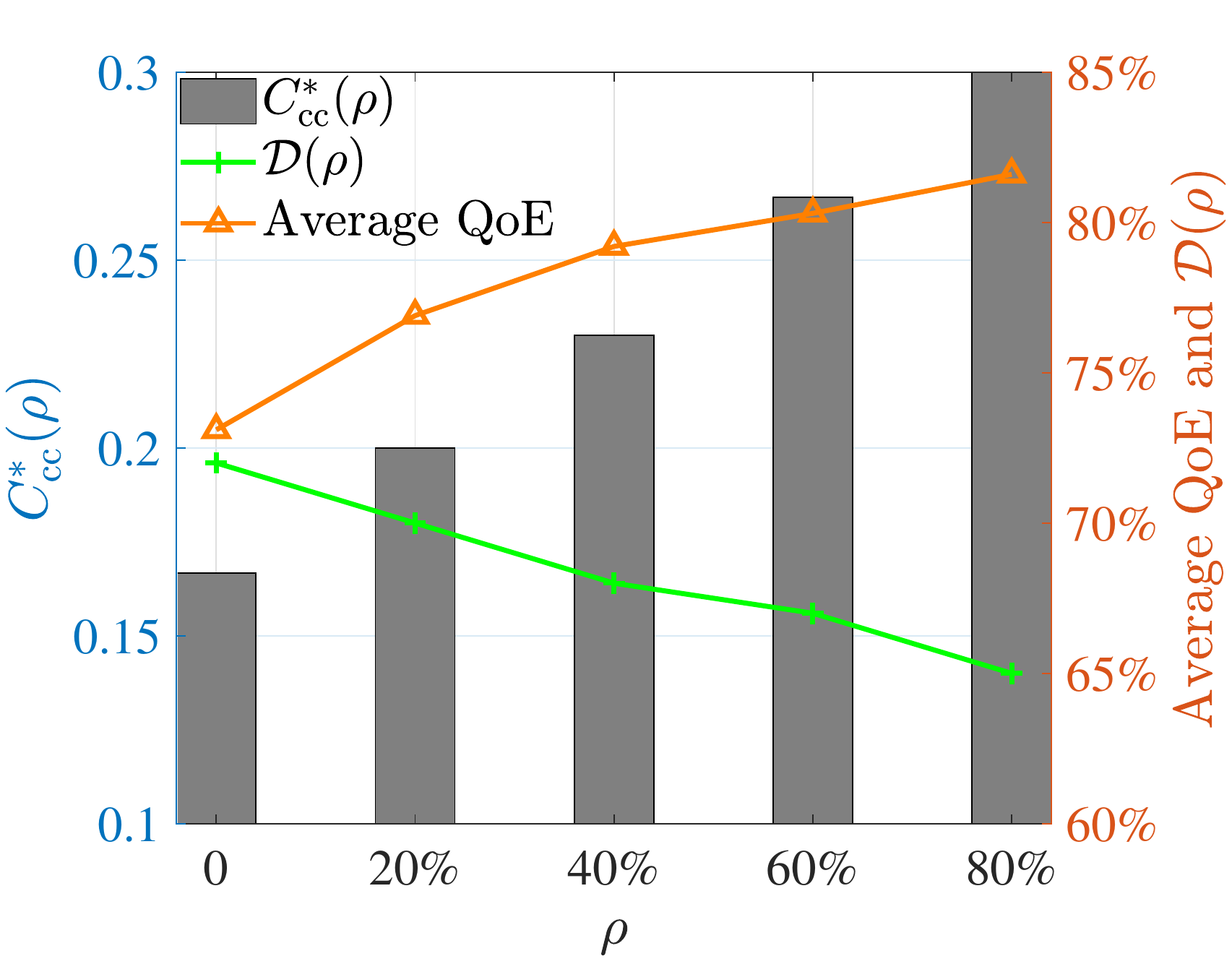}
		\end{minipage}
	}\\
	\subfloat[Tailored position-only predictor]{\label{Fig:qoe_rho_C_cc_Pos_Only_T_cc_m_6}
		\begin{minipage}[c]{0.85\linewidth}
			\centering
			\includegraphics[width=1\textwidth]{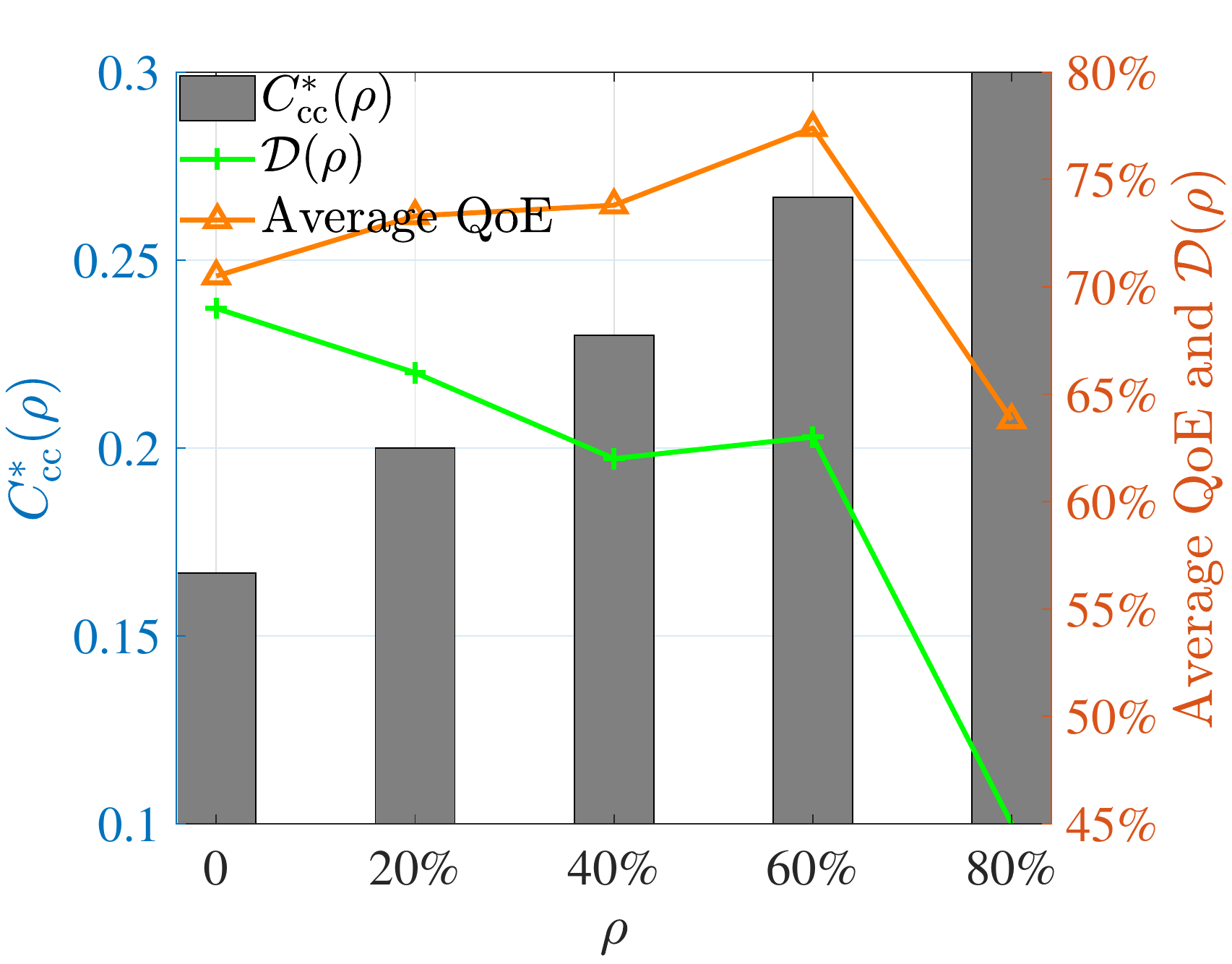}
		\end{minipage}
	}
	\caption{Average QoE, CC capability, and DoO v.s. DoP, $1/T_{\mathrm{cc}}^M=0.17$.}\label{Fig:QoE, CC capability, and DoO v.s. DoP}	
\end{figure}

\section{Conclusion}
In this paper, we investigated the impact of privacy on the VR video streaming system. We first collected a privacy dataset, verified that users indeed have privacy requirements. Then, we employed duration optimization and analysis of the obtained closed-form solution as tools, to find the relation between DoP and QoE. When DoP is considered, a predictor with variable length of input and output is more suitable.
We found a resource-saturated region where DoP has no impact and a resource-unsaturated region where DoP has contradictory impact on the QoE. The increase of DoP on the one hand improves the CC capability, on the other hand, it degrades the average-DoO. The overall impact depends on which factor the QoE is dominated. Simulation with two predictors validated the analysis and showed that the impact of DoP on QoE can also be effected by other influence factors, e.g., the robustness of a predictor to the variation of DoP and the assigned resource rates.

\bibliographystyle{unsrt}
\bibliography{acm_mm}


\end{document}